# Nowruz, Umar Khayyam, Calendar and Constellations

**Rizoi Bakhromzod**

S.U. Umarov Physical and Technical Institute of the National Academy of Sciences of Tajikistan

E-mail: rizo@physics.msu.ru

***Abstract.*** This article provides information about the factors that necessitated the need to adjust the calendar during the time of Umar Khayyam, the method of choosing the beginning of the year, and the length of the months. It also attempts to correct the common error associated with "the arrival of the Sun in Aries on Nowruz."

One of the great contributions of Umar Khayyam, which demonstrates his genius and profound astronomical knowledge, is the correction of the existing solar calendar, which later became known as the Jalali Calendar. This article attempts to evaluate Khayyam's contributions and analyzes his choice of the vernal equinox as the start of the year. The vernal equinox is the point in the sky where the center of the Sun's disk, during its apparent motion, crosses from the southern celestial hemisphere to the northern hemisphere. The depiction of the vernal equinox for our time, that is, the year 2024, is presented in Figure 1.

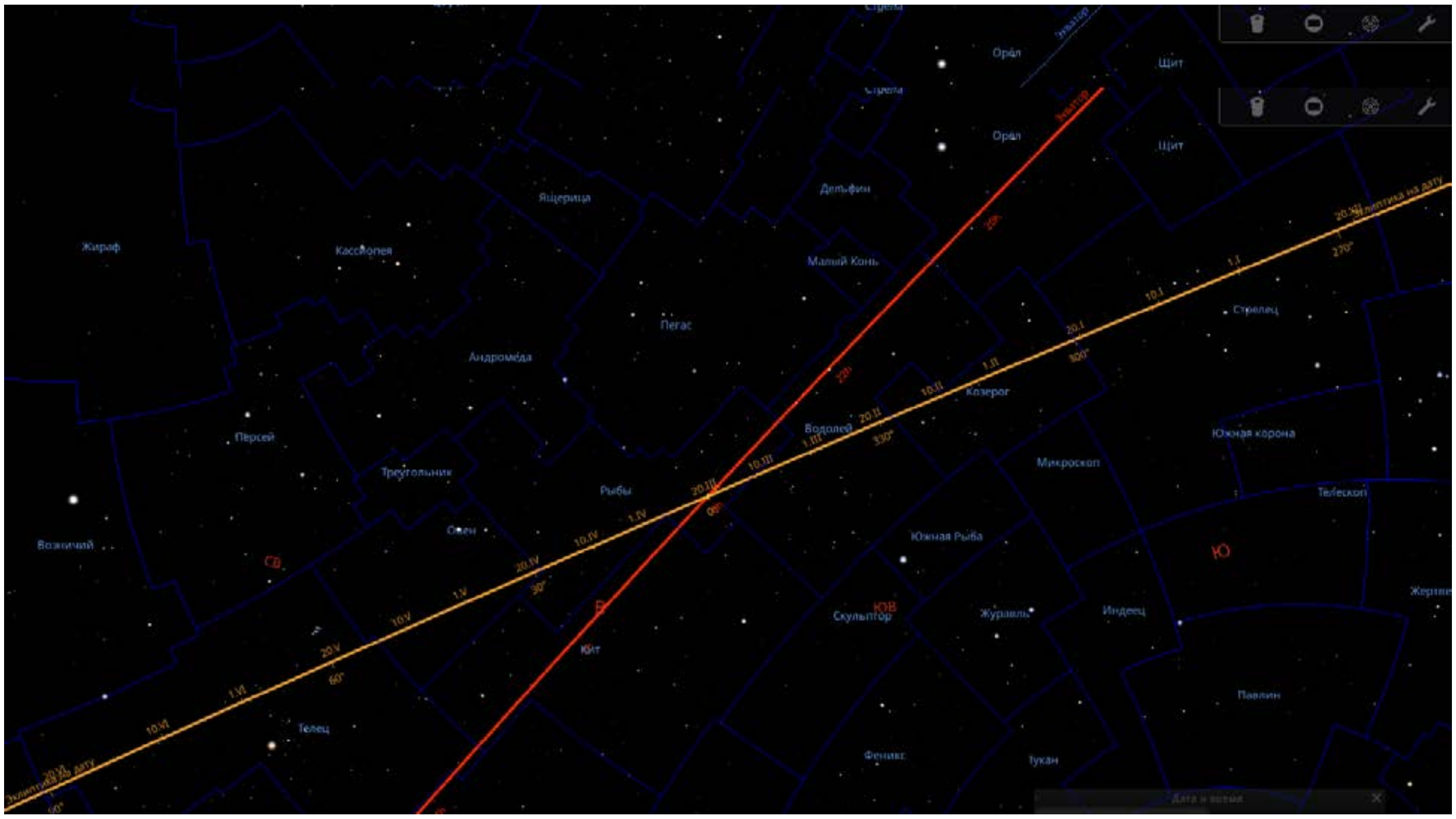

**Figure 1**. The point of the vernal equinox that the Sun crossed at 08:06:21 on March 20, 2024. The celestial equator is shown in red, and the ecliptic is shown in yellow.

The study should begin with an understanding of the calendar in the history of the Iranian people, as the existing calendar before Khayyam's time served as the basis for corrections [1]. It is estimated that before the Achaemenid period, the Iranian people used a lunar calendar of 6 months, each consisting of approximately two full lunar months. Later, to align with the Babylonian calendar, they used 12 synodic months, each containing 29 or 30 days. To match the tropical year, an additional month of 30 days was added every six years [2]. The length of the added month has been a topic of debate among scholars, but in ancient Iranian texts, such as the "Bundahishn," varying lengths of the year can be found. For instance, in chapter 5, the length of the year is stated to be 365 days and 5 hours and several minutes, whereas in chapter 25, it is emphasized that the length of the year is "one solar cycle from Aries to the end of the months," which amounted to 365 days and 6 hours and several minutes [3].

During the Achaemenid period, at least until 459 BC, a combined lunar and solar calendar was used. Between 471 and 401 BC, the Babylonian calendar was also used in Aramaic documents issued by the Persian government. According to the Roman historian Quintus Curtius Rufus, who wrote about 333 BC, the Persian year consisted of 360 days [4].

The calculation of the calendar in the Achaemenid era began, as in Babylonian tradition, with the ascension of a new king to the throne, and this practice continued during the Sassanian era. According to Biruni, the Sassanian calendar was solar, consisting of 365 days, and to align with the solar cycle, an extra month was added once every 120 years. The Khwarizmians and Sogdians, who were part of the Sassanian state, also used this calendar [5]. The last pre-Islamic Persian calendar started from the ascension of the last Sassanian king, Yazdegerd III, which corresponded to June 16, 632 AD [6, 7]. This calendar, known as the Yazdegerd calendar, remained unchanged for about two and a half centuries even after the Arab conquest of the Sassanian state and continued to be used in the everyday life of Zoroastrians [8].

Since the solar calendar was linked to the seasons, it was suitable for both Zoroastrian religious ceremonies and agricultural activities, such as sowing and harvesting. In the early stages of their conquests, the Arabs did not interfere with the calendars of the local people. However, the mismatch between the solar and lunar calendars created difficulties in collecting taxes. Biruni mentions in "Chronology of Ancient Nations" that Caliph Umar ibn al-Khattab considered using a calendar but ultimately decided against using the Persian calendar, thus adopting the Hijri lunar calendar. This problem resurfaced during the reign of Caliph al-Mutawakkil when he demanded taxes before the grain harvest was ready, which led to public discontent. When al-Mutawakkil summoned a Zoroastrian priest to resolve the issue by using the Persian calendar, it was discovered that due to the lack of leap months for nearly two and a half centuries, Nowruz had shifted approximately two months earlier than its original time during the Sassanian era. Al-Mutawakkil ordered the correct time of Nowruz to be determined and instructed that taxes be collected accordingly. However, he was killed before this could be fully implemented. The reform of the calendar was completed by Caliph al-Mu'tadid [9], adding two leap months to the Yazdegerd year. Nowruz (the first day of the new year) was moved from Saturday, the 1st of Farvardin (April 12), to Wednesday, the 1st of Khordad (May 12), to facilitate tax collection. Thus, a new calendar known as the Khiraji calendar, consisting of 365 days, 5 hours, 46 minutes, and 24 seconds, was introduced, starting from March 19, 611 AD [3, 10].

However, the Khiraji calendar did not become widely adopted, and the issue of timekeeping persisted until the reign of Sultan Jalal al-Din Malikshah Seljuk (1055-1092). Additionally, the celebration of Nowruz had shifted to the middle of Pisces due to the lack of leap years, and the vernal equinox occurred on the 19th of Farvardin. Recognizing this issue, the learned vizier of the Seljuks, Abu Ali Hasan ibn Ali ibn Ishaq al-Tusi, known as Nizam al-Mulk, proposed to the Sultan that the calendar be revised. Following the Sultan's order, a group of prominent astronomers and mathematicians was established. According to Umar Khayyam himself, "The sages of the era were brought from Khorasan, and every necessary instrument, including walls and astrolabes, was made, and Nowruz was brought to Farvardin." According to various sources, this group, led by Khayyam, included scholars such as Abdurrahman al-Khazini, Abu'l-Abbas Fazl ibn Muhammad al-Lukari, Abdul-Muzaffar al-Isfizari, Maymun ibn Najib al-Wasiti, and Ahmad Ma'ruf al-Bayhaqi [11, 12]. However, some researchers have questioned al-Khazini's participation in this group.

To accurately calculate time and adjust the calendar, precise observations of the Sun's movement against the background of constellations and the determination of the vernal equinox were necessary. Therefore, under Khayyam's leadership, these scholars established the Isfahan Observatory, which operated from 1074 to 1092, ceasing its activity after the death of Sultan Malikshah [13]. After five years of observations, the astronomers proposed a new calendar to the Sultan, which began on Friday, the 9th of Ramadan in the year 471 AH (according to some sources, the 10th of Ramadan), corresponding to March 15, 1079 Julian (March 21, 1079 Gregorian). According to the Hijri solar calendar, instead of the 19th of Farvardin in the year 458, the first of Farvardin was adopted, correcting the drift of Nowruz [14]. This day corresponds to the 19th of Farvardin in the year 448 Yazdegerdi and the 19th of Farvardin in the year 468 Khiraji. The important distinction of this calendar from others was that the beginning of the year was linked to the vernal equinox (i.e., the moment when the center of the Sun reaches the equinox point), and the remaining months were calculated based on the Sun's passage from one constellation to another [14, 15].

The new calendar was named the Jalali Calendar after the Sultan, also known as the Maliki or Malikshahi Calendar. According to various sources, it had a 33-year cycle with 8 leap years. In this calendar, the months of the first half of the year consisted of 31 days, while the remaining months had 30 days, and the last month had 29 days in a regular year [16]. In addition to this calculation, researchers have also presented other sequences of leap years.

## The Precessional Movement

It is important to note here that Khayyam's and his colleagues' decision to choose the vernal equinox as the start of the year and the day of Nowruz is both fascinating and historically significant. Due to the precession of the equinoxes, which is caused by the Earth's precessional motion, the points of the vernal and autumnal equinoxes shift each year in the direction of the Sun's apparent annual motion by approximately 50.28796 arcseconds [17]. This means that in each sidereal year, Nowruz occurs about 20 minutes and 24 seconds earlier than in the previous year. According to calculations, the vernal equinox point returns to its original position after 25,771,575.34 years, completing a full cycle.

According to Ptolemy (Claudius Ptolemy), in the 8th book of his work "Almagest," the discoverer of the precessional movement was Hipparchus (190-120 BC). Ptolemy mentions the lost book of Hipparchus and writes that Hipparchus compared his observational results with those of Timocharis (320-260 BC) and Aristillus (~280 BC) and concluded that the longitude of the star Spica in the constellation Virgo had shifted by approximately 2° relative to the vernal equinox point. As a result, he concluded that the rate of drift of the equinox point was approximately 1° per century, completing a full cycle in 36,000 years [18].

Among Muslim scholars, Al-Battani, in his work "Al-Zij al-Sabi," determined the movement of the vernal and autumnal equinox points towards the Sun's annual path to be 54.5 arcseconds per year, or 1° in 66 years [19]. The famous Tajik astronomer Abd al-Rahman al-Sufi (903-986) wrote in his book "Book of Fixed Stars": "And all people agree that fixed stars have a movement towards the ecliptic; however, Ptolemy's opinion is that it moves 1 degree every hundred years, while the view of the 'tested ones' and those who came after Ptolemy is that it moves 1 degree every sixty-six years." Here, the term 'tested ones' refers to a group of astronomers who compiled "Al-Zij al-Ma'muni for the Tested" in Baghdad under the leadership of Yahya ibn Abu Mansur during the years 829-830 for Caliph al-Ma'mun [20]. The displacement of the vernal equinox point against the backdrop of the starry sky over the past 6,000 years is shown in Figure 2.

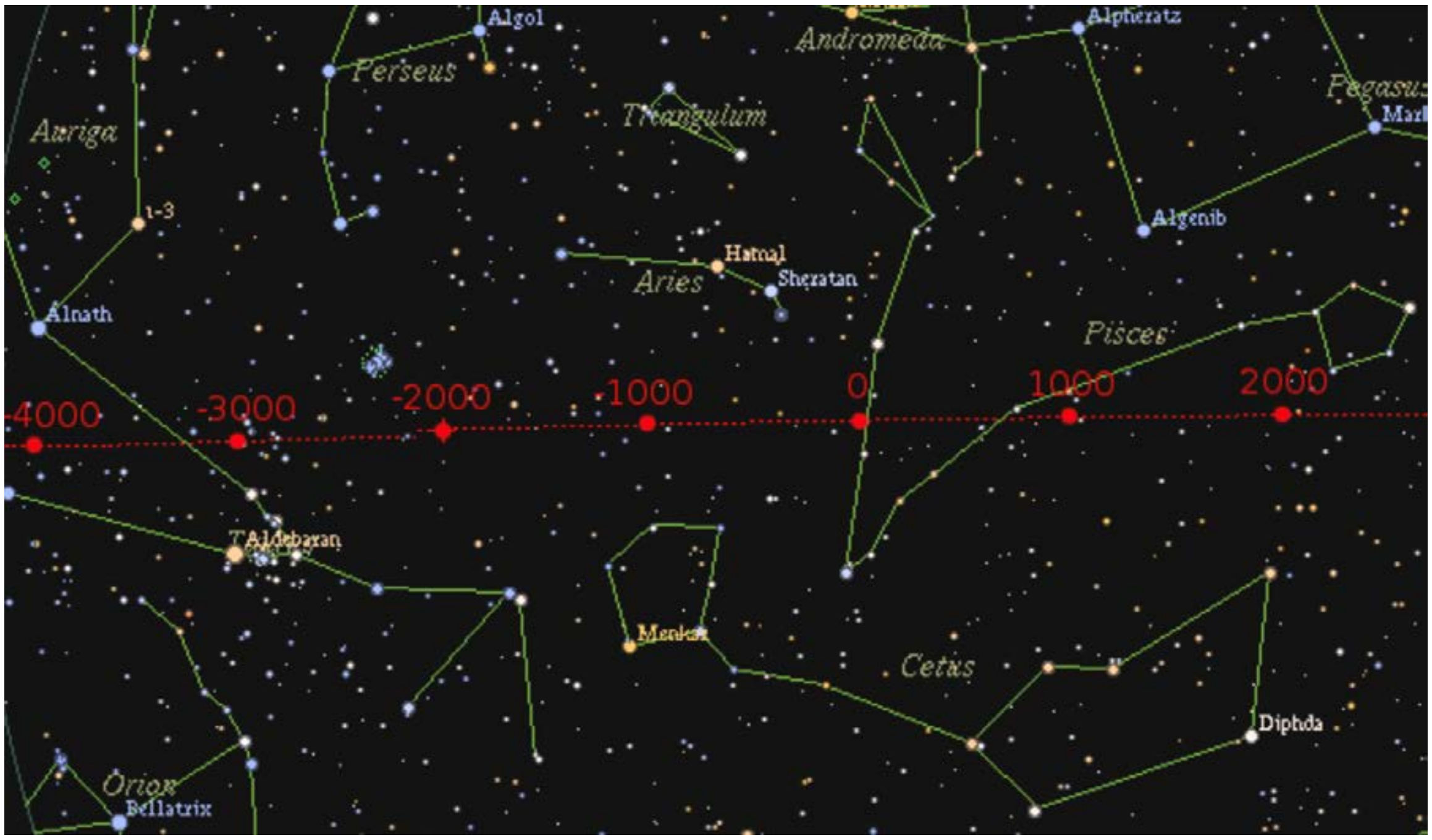

**Figure 2**. Displacement of the vernal equinox point against the backdrop of the starry sky over the past 6,000 years.

## The Division of the Sky into Constellations

In ancient times, to facilitate learning about the sky and systematizing the positions of stars, people connected the stars into recognizable shapes and named them, which are collectively called constellations [21]. The shapes formed by connecting stars differed across cultures and civilizations, and people used them to tell stories, myths, and beliefs. Constellations were also used for navigation and determining time. It is not precisely known when the division of stars into constellations began, but their astronomical (in this context, astrological) use started in Babylon [22], later spreading to Ancient Greece and other countries.

The first complete work describing constellations and explaining the 48 known constellations of its time was Ptolemy's "Almagest." Another significant work is Abd al-Rahman al-Sufi's "Book of Fixed Stars," which elaborates on "Almagest." Al-Sufi sought to blend the knowledge of the Greeks with the concepts of Eastern cultures, providing new data on the coordinates of stars and correcting Ptolemy's mistakes. He even added 132 new stars to his star catalog [23].

While in Ptolemy's and al-Sufi's era, the celestial sphere was divided into 48 constellations or 'kawakib,' in 1922, the International Astronomical Union approved the names of 88 constellations, and in 1930, their boundaries were defined [24, 25]. Until 1930, the boundaries of constellations were vague and varied across different cultures, but after that, they were unified and remain unchanged to this day.

Among the 88 constellations, 12 are historically more important and well-known; they are called the zodiac constellations, located along the zodiacal belt. In ancient times, for precise month calculations, the apparent path of the Sun was divided into 12 equal parts (each 30°), and the name of each constellation matched the position of the celestial figure within its boundaries. This division was approximately made in the 5th century BC by Babylonian astronomers, with each month corresponding to one constellation [26]. However, this division is abstract because the sizes of constellations in the sky are not equal, and the Sun does not remain in each constellation for the same duration. Therefore, we must acknowledge that there are two ways of understanding constellations: (1) historical abstract constellations used only in astrology today, with months named after them, and (2) constellations as defined sectors of the sky approved by the International Astronomical Union. It is important to note that in ancient Babylon, both types of constellations were not distinguished from each other. Table 1 shows the names of the zodiac constellations in different languages.

**Table 1. Names of the Zodiac Constellations and Their Corresponding Months in the Iranian Calendar**

| № | Arabic | Persian (Constellation Name) | Persian (Month Name) | Russian |
|---|---|---|---|---|
| **1** | Hamal Kabsh | Barreh | Farvardin | Oven |
| **2** | Sawr | Gav | Ordibehesht | Telets |
| **3** | Jawza Taw'amun | Do-pakar | Khordad | Bliznetsy |
| **4** | Saratan | Qarzang | Tir | Rak |
| **5** | Asad | Shir | Mordad | Lev |
| **6** | Sunbula Uzra | Khosha | Shahrivar | Deva |
| **7** | Mizan | Tarazu | Mehr | Vesy |
| **8** | Aqrab | Kajdum | Aban | Skorpion |
| **9** | Qaws Rami | Nimasb | Azar | Strelets |
| **10** | Jadi | Behi | Dey | Kozerog |
| **11** | Dalv | Dul | Bahman | Vodoley |
| **12** | Hut | Mahi | Esfand | Ryby |

Analyzing these two facts—precessional movement and constellations—we see that due to precession, the equinox points gradually move from one constellation to another over long periods. For example, while the vernal equinox was in Aries, it is now in Pisces.

Abd al-Rahman al-Sufi understood this well and wrote: "... Since these forms on the Zodiac Belt existed more than three thousand years ago in divisions other than these, there should be nothing strange if, at that time, the names of divisions matched the images there—that is, when Aries was in the twelfth division, Taurus was in the first, and the divisions were named accordingly. During the times of Timocharis and before him, when they made a new determination of arcs, they found that Aries was in the first division after the intersection of the celestial equator and the ecliptic, and thus named it the first division as Aries, the second as Taurus, and the third as Gemini" [20]. In other words, al-Sufi recognized that approximately 3,000 years before his time, the vernal equinox point was located in Taurus. If we consider the present boundaries of constellations, the transition of the vernal equinox from Taurus to Aries occurred around 1860-70 BC, and in the time of Timocharis, the first constellation after the vernal equinox was Aries, hence starting the calendar from this constellation. However, by the 1st century AD, the vernal equinox had already shifted to Pisces, yet al-Sufi, following Ptolemy, began naming the twelve zodiacal constellations from Aries, using the names of the celestial figures present in them.

Since the source of astronomical knowledge for our ancestors was Babylonian and Greek works, most of our scholars considered the beginning of the year and the arrival of Nowruz as coinciding with the Sun entering Aries. For example, the great Firdausi wrote:

*"When the Sun entered Aries,*
*The world became splendid, orderly, and radiant."*

Or Umar Khayyam himself wrote in his "Nowruznameh": "The reason for naming Nowruz was that they knew the Sun had two cycles: one of 365 days and a quarter, returning to the first minute of Aries at the same time and day as before... And when Jamshid discovered that day, he named it 'Nowruz' and established a celebration." Biruni, in "Kitab al-Tafhim," also wrote about this day: "It is the first day of Farvardin (Aries), and thus it was called the New Day because it is the beginning of the new year." In "Chronology of Ancient Nations," he states: "In our time (Nowruz) coincides with the Sun's entry into Aries, which marks the beginning of spring." Nasir al-Din al-Tusi, in his "Ilkhanic Tables," wrote about Khayyam's calendar: "It is established that the beginning of the year is the day the Sun enters Aries, marking the true spring. At the beginning of each month, the Sun enters the zodiac constellation corresponding to that month." This idea is also repeated in the "New Gurkanian Tables" [14].

However, the idea that the Sun's entry into Aries has remained unchanged until our time has been overlooked by modern researchers, who continue to repeat this error, linking the arrival of Nowruz to Aries. For instance, the renowned Tajik researcher Muhammadqul Hazratqulov, in his book "Nowruz: The World-Illuminating Celebration and Other Traditional Holidays of the Year," while explaining the "Astronomical Foundations of Nowruz" based on Iranian scholars, correctly describes precession but mistakenly associates the beginning of Khayyam's calendar not with the vernal equinox point but rather correlates the vernal equinox with the start of Farvardin [27]. Similarly, the well-known Tajik scholar Akbari Turson, in his book "The Universe and the Flow of Human Knowledge," provides information about precession [28], but in another article writes: "Nowruz is a festival of the vernal equinox, which is a purely natural event and is equivalent to another purely natural event—the Sun entering Aries during its annual journey along the zodiac belt" [29]. The Iranian scholar Reza Sha'bani also repeats this misconception in his book [30]. Tajik philosophers [31], physicists [32], and astronomers [33, 34] have also echoed this incorrect idea of the Sun entering Aries on Nowruz. This misconception is often heard in the media and on public platforms, especially on television [36-38]. However, in modern times, the Sun enters Aries on April 19.

The great contribution of Khayyam lies precisely in the fact that he linked the start of the calendar not to the Sun's entry into Aries but to the vernal equinox point, placing the months in sequence after it. The names of the months in Khayyam's calendar are traditional and inherited from previous calendars, but they do not have a direct relationship to celestial constellations;

they are symbolic. It is important to understand that Aries or Pisces in Khayyam's calendar are simply the names of months.

Furthermore, contrary to the ancient beliefs—especially astrological notions that still exist—the Sun in its apparent annual movement does not pass through 12 constellations but rather through 13. The Sun's passage through Ophiuchus is not considered in the works of our predecessors. Additionally, the duration of the Sun's stay in each constellation is not equal, depending on the constellation's size, ranging from 7 days (in Scorpio) to 45 days (in Virgo) (Figure 3).

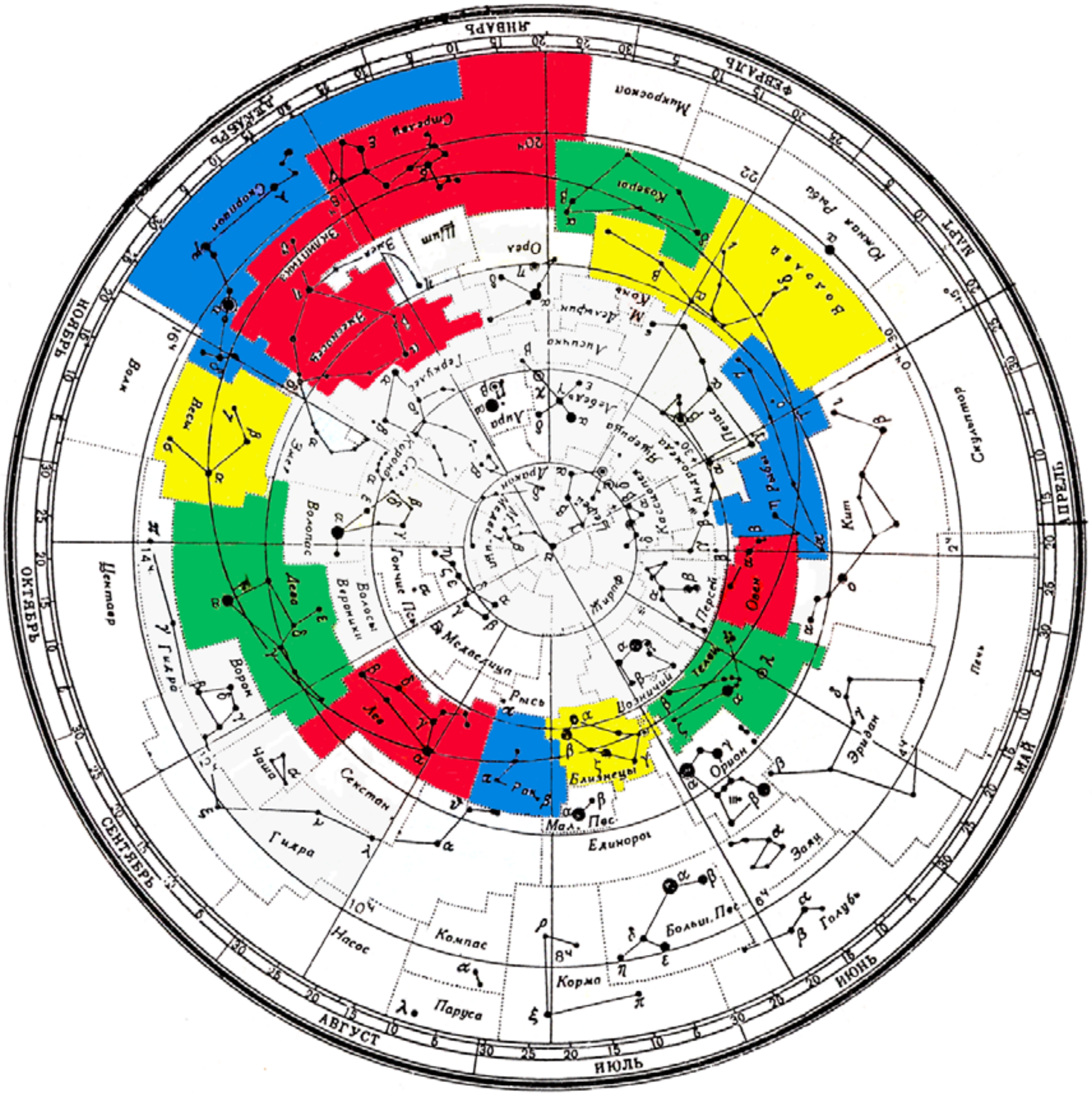

**Figure 3**. Map of the starry sky showing the ecliptic line and the constellations that the Sun passes through.

The correlation between Khayyam's calendar months and the months of the Gregorian calendar, the Sun's positions in the constellations, the duration of its stay, and the period during which the vernal equinox point is located in constellations are presented in Table 2 [39].

**Table 2. Names of Khayyam's Calendar Months, Astrological and Astronomical Dates, and the Sun's Position in the Constellations**

| Month Name | Astrological Dates | Constellation | Astronomical Dates | Duration of the Sun's Stay | Position of the Vernal Equinox in Constellations |
|---|---|---|---|---|---|
| **Farvardin** | March 21 - April 20 | Aries | April 19 - May 13 | 25 | 2000 BC - 100 BC |
| **Ordibehesht** | April 21 - May 21 | Taurus | May 14 - June 19 | 37 | 4500 BC - 2000 BC |
| **Khordad** | May 22 - June 21 | Gemini | June 20 - July 20 | 31 | 6600 BC - 4500 BC |
| **Tir** | June 22 - July 22 | Cancer | July 21 - August 9 | 20 | 8100 BC - 6600 BC |
| **Mordad** | July 23 - August 23 | Leo | August 10 - September 15 | 37 | 10800 BC - 8100 BC |
| **Shahrivar** | August 24 - September 23 | Virgo | September 16 - October 30 | 45 | 12000 AD - 15300 AD |
| **Mehr** | September 24 - October 23 | Libra | October 31 - November 22 | 23 | 10300 AD - 12000 AD |
| **Aban** | October 24 - November 22 | Scorpio | November 23 - November 29 | 7 | 8600 AD - 10300 AD |
| **Havvo Ophiuchus** | Not recognized by astrologers | Ophiuchus | November 30 - December 17 | 18 | — |
| **Azar** | November 23 - December 21 | Sagittarius | December 18 - January 18 | 32 | 6300 AD - 8600 AD |
| **Dey** | December 22 - January 20 | Capricorn | January 19 - February 15 | 28 | 4400 AD - 6300 AD |
| **Bahman** | January 21 - February 19 | Aquarius | February 16 - March 11 | 24 | 2700 AD - 4400 AD |
| **Esfand** | February 20 - March 20 | Pisces | March 12 - April 18 | 38 | 100 BC - 2700 AD |

Considering all this information, we can confidently state that the months in Khayyam's calendar should be understood as mere month names,

without associating them directly with the constellations. This approach helps to avoid further mistakes by researchers.

## Bibliography

1. Panaino A. "Pre-Islamic Calendars." Encyclopaedia Iranica. 1990, Vol. 4, pp. 658-668.
2. Taqizadeh S. H. "The Old Iranian Calendars Again." Bulletin of the School of Oriental and African Studies, 1952, Vol. 14, No. 3, pp. 603-611.
3. Akrami M. "The Development of the Iranian Calendar: Historical and Astronomical Foundations." arXiv preprint arXiv:1111.4926. 2011.
4. Bickerman E. J. "The Zoroastrian Calendar." Archiv Orientální. 1967, Vol. 35, pp. 197-207.
5. Abu Rayhan al-Biruni. "Chronology of Ancient Nations." Dushanbe: Irfon, 1990, 432 pages.
6. Tyler-Smith S. "Coinage in the Name of Yazdgerd III (AD 632-651) and the Arab Conquest of Iran." The Numismatic Chronicle (1966-), 2000, Vol. 160, pp. 135-170.
7. Cristoforetti S. et al. "On the Era of Yazdegard III and the Cycles of the Iranian Solar Calendar." Annali di Ca'Foscari. Serie Orientale, 2014, Vol. 50, No. 1, pp. 159-172.
8. Dalvand H. "The Reform of the Yazdgerdi Calendar in the Year 375 Yazdgerdi According to the Rivaayat of Farnbag-SRosh." Iranian Journal for the History of Islamic Civilization, 2021, Vol. 53, No. 2, pp. 399-417.

9. Taqizadeh S. H. "Various Eras and Calendars Used in the Countries of Islam." Bulletin of the School of Oriental and African Studies, 1939, Vol. 9, No. 4, pp. 903-922.
10. Rezā ʻAbdullāhī. "Calendars ii. In Islamic Period." Encyclopedia Iranica, online version: http://www.iranicaonline.org/articles/calendars.
11. Umar Khayyam. "Nowruznameh." Dushanbe: Adib, 2012, 352 pages.
12. Heydari-Malayeri M. "A Concise Review of the Iranian Calendar." arXiv preprint astro-ph/0409620, 2004.
13. Sayılı A. "The Observatory in Islam and its Place in the General History of the Observatory." Türk Tarih Kurumu yayınlarından, 1960.
14. Rosenfeld B. A., Yushkevich A. P. "Omar Khayyam." Academy of Sciences of the USSR. Moscow: Nauka, 1965, 192 pages.
15. Thomann J. "The Institution of the Jalālī Calendar in 1079 CE and Its Cohabitation with the Older Persian Calendar." In: Calendars in the Making: The Origins of Calendars from the Roman Empire to the Later Middle Ages, Brill, 2021, pp. 210-244.
16. Klimishin I. A. "Calendar and Chronology." 3rd ed., revised and enlarged. Moscow: Nauka, Main Editorial Office for Physical and Mathematical Literature, 1990, 480 pages.
17. Capitaine N., Wallace P. T., Chapront J. "Expressions for IAU 2000 Precession Quantities." Astronomy & Astrophysics, 2003, Vol. 412, No. 2, pp. 567-586.
18. Ptolemy Claudius. "Almagest: A Mathematical Composition in Thirteen Books." Translated from Ancient Greek by I. N. Veselovsky. Moscow: Nauka, 1998, 672 pages.

19. Angelo J. A. "Encyclopedia of Space and Astronomy." Infobase Publishing, 2014.
20. Abd al-Rahman al-Sufi. "Book of Fixed Stars." Dushanbe: Donish, 2020, 361 pages.
21. Kononovich E. V., Moroz V. I. "General Course of Astronomy: Textbook." Edited by V. V. Ivanov. Moscow: Editorial URSS, 2004, 544 pages.
22. Britton J. P. "Studies in Babylonian Lunar Theory: Part III. The Introduction of the Uniform Zodiac." Archive for History of Exact Sciences, 2010, Vol. 64, No. 6, pp. 617-663.
23. Hafez I., Stephenson F. R., Orchiston W. "The Investigation of Stars, Star Clusters, and Nebulae in 'Abd al-Rahman al-Sufi's Book of Fixed Stars." In: New Insights From Recent Studies in Historical Astronomy: Following in the Footsteps of F. Richard Stephenson. A Meeting to Honor F. Richard Stephenson on His 70th Birthday. Springer, 2014, Vol. 43, pp. 143.
24. Schlesinger M. F., Bosler M. M., Chant et al. "Commission on Notation of Units and Publishing Economy." Transactions of the International Astronomical Union, Vol. 4, 1933, pp. 19.
25. Delporte E. "Scientific Delimitation of Constellations (Tables and Maps)." Cambridge, 1930.
26. Ossendrijver M. "Mesopotamian Science." In: The Encyclopedia of Ancient History, 2013.
27. Muhammadqul Hazratqulov. "Nowruz: The World-Illuminating Celebration and Other Traditional Holidays of the Year." Dushanbe: Er-Graf, 2012, 484 pages.

28. Tursonov A. "The Universe and the Flow of Human Knowledge." Dushanbe: Irfon, 1973, 120 pages.
29. Turson A. "Spring Unveils Colors and Secrets (On the Connection Between the Nowruz Celebration and Religion)." Journal "Asia and Europe," 2019, No. 1, pp. 20-41.
30. Sha'bani R. "Traditions and Customs of Nowruz." Dushanbe: Payvand, 2008, 235 pages.
31. Jangibekova S. "Mehrgon: Yesterday and Today." Available at: https://ifppanrt.tj/tj/ilm-va-navidho/andeshai-muhakkikon/item/1358-mehrgon-dir-z-va-imruz.html.
32. Qurbanov N. "Nowruz—The Celebration of the Unity of Nature and Cosmos." Available at: https://ravshanfikr.tj/shinokhti-masoili-i-timo-va-sijos/Nowruz-ashni-ambastagii-tabiatu-kayhon.html.
33. Khujanazarov H.F., Pirova V.S. "Astronomical Nowruz or the Beginning of the New Year." Available at: https://ravshanfikr.tj/shinokhti-masoili-i-timo-va-sijos/Nowruzi-astronom-jo-ibtidoi-ruzu-soli-nav.html.
34. Rahmatulloyeva F. "What is Astronomical Spring?" Sadoi Mardum, 2012.
35. "Nowruz in Tajikistan: How People Celebrate in Badakhshan, Sughd, and Khatlon." Available at: https://halva.tj/tj/articles/education/Nowruz_dar_tojikiston_mardumi_badakhshonu_sugd_va_khatlon_onro_ch_guna_ashn_megirand/.
36. "If Nowruz Falls on a Friday..." Available at: https://farazh.tj/sahifai-asosy/farhang/agar-Nowruz-ruzi-juma-buvad/.
37. Nazarzoda S. "Our Ancestors' Nowruz." Izar.tj, March 15, 2012.

38. Usmon Nazir. "A Look at the Nowruz Celebration." Sadoi Mardum, March 15, 2012, No. 31 (2858), p. 1.
39. Kaler J. B. "The Ever-Changing Sky: A Guide to the Celestial Sphere." Cambridge University Press, 2002.